\documentclass[fleqn,usenatbib]{mnras}

\usepackage{newtxtext,newtxmath}
\usepackage[T1]{fontenc}

\DeclareRobustCommand{\VAN}[3]{#2}
\let\VANthebibliography\thebibliography
\def\thebibliography{\DeclareRobustCommand{\VAN}[3]{##3}\VANthebibliography}

\usepackage{graphicx}	% Including figure files
\usepackage{amsmath}	% Advanced maths commands
\usepackage{caption}
\usepackage[para, flushleft]{threeparttable}
\title[Discovering Strongly-lensed QSOs From Unresolved Light Curves]{Discovering Strongly-lensed QSOs From Unresolved Light Curves}

\author[Shu et al. 2020]{
Yiping Shu,$^{1}$\thanks{E-mail: yiping.shu@mpa-garching.mpg.de}
Vasily Belokurov,$^{1}$
and N. Wyn Evans$^{1}$
\\
$^{1}$Institute of Astronomy, University of Cambridge, Madingley Road, Cambridge, CB3 0HA, UK\\
}

\date{Accepted XXX. Received YYY; in original form ZZZ}

\pubyear{2020}

\begin{document}
\label{firstpage}
\pagerange{\pageref{firstpage}--\pageref{lastpage}}
\maketitle

\begin{abstract}
We present a new method of discovering galaxy-scale, strongly lensed QSO systems from unresolved light curves using the autocorrelation function. The method is tested on five rungs of simulated light curves from the Time Delay Challenge 1 that were designed to match the light-curve qualities from existing, ongoing, and forthcoming time-domain surveys such as the Medium Deep Survey of the Panoramic Survey Telescope And Rapid Response System 1, the Zwicky Transient Facility, and the Rubin Observatory Legacy Survey of Space and Time. Among simulated lens systems for which time delays can be successfully measured by current best algorithms, our method achieves an overall true positive rate of 28--58 per cent for doubly-imaged QSOs (doubles) and 36--60 per cent for quadruply-imaged QSOs (quads) while maintains $\lesssim$10 per cent false positive rates. We also apply the method to observed light curves of 22 known strongly lensed QSOs, and recover 20 per cent of doubles and 25 per cent of quads. The tests demonstrate the capability of our method for discovering strongly lensed QSOs from major time domain surveys. The performance of our method can be further improved by analysing multi-filter light curves and supplementing with morphological, colour, and/or astrometric constraints. More importantly, our method is particularly useful for discovering small-separation strongly lensed QSOs, complementary to traditional imaging-based methods. 
\end{abstract}


%%%%%%%%%%%%%%%%%%%%%%%%%%%%%%%%%%%%%%%%%%%%%%%%%%

%%%%%%%%%%%%%%%%% BODY OF PAPER %%%%%%%%%%%%%%%%%%

\section{Introduction}

Strongly lensed QSO systems have been shown to be a powerful tool for a broad range of important topics in astronomy and astrophysics. They can be used to measure the initial mass function (IMF) and dark-matter substructures in the lensing galaxies \citep[e.g.,][]{Mao98, Metcalf01, Dalal02, Pooley09, Nierenberg14, Oguri14, Schechter14, Jimenez-Vicente19}. They also serve as a unique probe of the size and temperature profile of black hole accretion discs and black hole/host galaxy scaling relations at cosmological distances \citep[e.g.,][]{Peng06, Ding17, Bate18, Morgan18, Cornachione20, Ding20}. The statistics of a well-defined sample of strongly lensed QSOs can provide a simple and robust constraint on the structure growth and hence dark energy \citep[e.g.,][]{Chae07, Oguri08, Oguri12}. Lastly, utilising time delays from strongly lensed QSOs, independent and precise measurements of the Hubble constant have been achieved \citep[e.g.,][]{Suyu13, Suyu14, Bonvin17, Wong20}, which are of particular importance given the growing tension in the Hubble constant measured by different approaches \citep[e.g.,][]{Addison18, Planck18, Riess19, Macaulay19} that could hint for new physics beyond the current standard cosmological model \citep[e.g.,][]{Zhao17, Renk17, DiValentino18}. 

Due to their intrinsic rarity, identification is a crucial step in strongly lensed QSO studies and applications. Discovering strongly lensed QSOs has traditionally relied on imaging and spectroscopic signatures that are a result of strong lensing, i.e. multiple QSOs with identical spectral profiles forming specific configurations. Almost all the $\approx$200 strongly lensed QSOs known to date have been discovered by imaging-based or spectroscopy-based methods \citep[e.g.,][]{Walsh79, Browne03, Oguri06, Jackson12, Inada12, More16, Lemon18, Shu18, Lemon19, Shu19, Chan20}. However, imaging and spectroscopy data alone, especially only in the optical, sometimes are insufficient for unambiguously distinguishing physically distinct binary QSOs from doubly-imaged QSOs when there is no observational evidence for a plausible deflector or extended lensing features of the QSO host. This is because binary QSOs can have very similar optical spectra due to the generic nature of QSO spectra, sometimes even more similar than doubly-imaged QSOs \citep[e.g.,][]{Peng99, Mortlock99}. 

Fortunately, strongly lensed QSOs have another distinct signature, which is lagged and so correlated variations between different lensed images of QSOs that show noticeable variability. Besides being used as an extra indicator of the lensing nature, this type of strong-lensing induced signature can be directly utilised as an additional discovery channel. For instance, \citet{Pindor05} introduced a method of identifying strongly lensed QSO pairs using a statistical metric computed from light curves of closely separated sources. \citet{Kochanek06} proposed an idea of selecting strongly lensed QSO candidates from difference images as spatially extended variable sources, which, by their estimations, should be dominated by strongly lensed QSOs at high Galactic latitudes ($\lvert b \rvert \gtrsim 20^{\circ}$). This idea has been more quantitatively validated and extended with real data \citep{Lacki09, Chao20b} or simulated data \citep{Chao20}. Recently, a colour- and variability-selected QSO candidate from the Optical Gravitational Lensing Experiment (OGLE) was found to have a neighbouring OGLE source with similar variability properties, and later spectroscopically confirmed to be a doubly-imaged QSO \citep{Kostrzewa-Rutkowska18}. 

In this work, we aim to develop a new method of discovering galaxy-scale, strongly lensed QSOs in the time domain, which can fully exploit existing time-domain surveys such as the Medium Deep Survey (MDS) of the Panoramic Survey Telescope And Rapid Response System 1 \citep[Pan-STARRS1,][]{PS1} and the Zwicky Transient Facility \citep[ZTF,][]{ZTF}, and will be of further importance in the upcoming era of the Rubin Observatory Legacy Survey of Space and Time \citep[LSST,][]{LSST09}. Our method is specifically designed to apply on joint light curves of the whole strongly lensed QSO systems. Compared to resolved/deblended light curves or difference images required by previous time-domain discovery methods mentioned above, joint light curves make the least demands on observing conditions across the entire survey and are the only product that does not require accurate point spread function modelling, which can be challenging and/or introduce artefacts that lead to performance degradation. Moreover, our method of using joint light curves is particularly capable of discovering strongly lensed QSOs with separations smaller than the resolutions of the time-domains surveys, which is highly complementary to traditional imaging-based methods. 

The rest of the paper is organised as follows. Section~\ref{sect:method} outlines the methodology. Tests on simulated and observed light curves are presented in Sections~\ref{sect:test_tdc1} and \ref{sect:test_real}. Discussions are given in Section~\ref{sect:discussions}, and Section~\ref{sect:conclusion} is the conclusion. 

\section{Methodology}
\label{sect:method}

Let $f(t_i)$ be the flux in a particular band of one QSO image in a strongly lensed QSO system at a series of time points $t_i$ ($i=$1, 2, 3, ..., N). The autocorrelation function (ACF) of time series $f(t_i)$ is defined as 
\begin{equation}
    \text{ACF } (t_{\rm lag}; f) = \frac{\langle [f(t_i)-\langle f(t_i) \rangle][f(t_i+t_{\rm lag})-\langle f(t_i) \rangle] \rangle}{\langle f^2(t_i) \rangle-\langle f(t_i) \rangle^2},
\end{equation}
where $\langle x \rangle$ denotes the mean of $x$. By this definition, the ACF of any data is one at zero time lag. The light curve in the same band of another lensed QSO image in the same system that has a relative time delay of $t_{\rm delay}$ and flux ratio of $\mathcal{R}$ is $f^{\prime} (t_i) \equiv \mathcal{R} f(t_i-t_{\rm delay})$.
The ACF of the joint light curve of these two images $F(t_i) = f(t_i) + f^{\prime} (t_i)$ is therefore
\begin{equation}
    \text{ACF} (t_{\rm lag}; F) = \frac{\langle [F(t_i) - \langle F(t_i) \rangle][F(t_i+t_{\rm lag}) - \langle F(t_i) \rangle] \rangle}{\langle F^2(t_i) \rangle - \langle F(t_i) \rangle^2}.
\end{equation}
Considering that $f(t_i)$ and $f(t_i-t_{\rm delay})$ are different portions of the same light curve with significant overlaps, we can make an approximation that the mean, standard deviation, and ACF of $f(t_i-t_{\rm delay})$ are equal to those of $f(t_i)$, and rewrite the ACF of the joint light curve as
\begin{align}
    \label{eq:ACF_joint}
    \text{ACF} (t_{\rm lag}; F) &= (1+\mathcal{R}^2) \times \text{ACF}(t_{\rm lag}; f) + \mathcal{R} \times \text{ACF}(t_{\rm lag}-t_{\rm delay}; f) \nonumber \\
    & + \mathcal{R} \times \text{ACF}(t_{\rm lag}+t_{\rm delay}; f).
\end{align}
We thus expect a positive excess in the ACF of the joint light curve around $t_{\rm lag} = t_{\rm delay}$. Similarly, a joint light curve of more than two lensed images will show ACF excesses at several $t_{\rm lag}$ that correspond to the time delays between different image pairs. As a result, ACF excesses can be used as an indicator of a strongly lensed QSO system.

%The numerator in the above equation can be decomposed into the sum of four terms, i.e. $\langle [f(t_i)-\langle f(t_i) \rangle][f(t_i+t_{\rm lag})-\langle f(t_i) \rangle] \rangle$, $\mathcal{R}^2 \langle [f(t_i-t_{\rm delay})-\langle f(t_i-t_{\rm delay}) \rangle][f(t_i-t_{\rm delay}+t_{\rm lag})-\langle f(t_i-t_{\rm delay}) \rangle] \rangle$, $\mathcal{R} \langle [f(t_i)-\langle f(t_i) \rangle][f(t_i-t_{\rm delay}+t_{\rm lag})-\langle f(t_i-t_{\rm delay}) \rangle] \rangle$, and $\mathcal{R} \langle [f(t_i+t_{\rm lag})-\langle f(t_i) \rangle][f(t_i-t_{\rm delay})-\langle f(t_i-t_{\rm delay}) \rangle] \rangle$. 

Through some tests, however, we find it technically difficult to effectively identify such excesses from the ACFs of the light curves while maintaining an acceptably low false positive rate (FPR). In fact, \citet{Geiger96} already discussed this difficulty when trying to measure time delays from ACFs of joint light curves. They instead developed a technique to reconstruct light curves of individual lensed QSO images by treating the time delay and flux ratio as free parameters, assuming the input joint light curve is from a strongly lensed QSO. However, they pointed out that the solutions of the reconstruction were generally not unique, and additional observational constraints (such as the true flux ratios from supplementary, resolved data) were required to narrow down the parameter space. Therefore, the reconstruction technique in \citet{Geiger96} is not suitable for the purpose of discovery. 

We develop a new method to tackle the above problem by analysing the residuals obtained from fitting a light curve with a sensible model that can describe the variability of single QSOs. That way, the ACF of the residual light curve for a single QSO should remain close to zero at any non-zero $t_{\rm lag}$, while we intuitively expect the ACF of the residuals from the joint light curve of two lensed QSO images to show excesses around $t_{\rm lag} = t_{\rm delay}$. Considering that current and forthcoming large-scale time-domain surveys are mainly in the optical, we demonstrate our method using the damped random walk (DRW) model that has been found to provide good descriptions to the long-term (above a few months) optical variability of QSOs \citep[e.g.,][]{Kelly09, Kozlowski10, MacLeod10, Zu13}. Nevertheless, our method is by no means limited to optical light curves and should work with any other model that can describe QSO variability (see Section~\ref{sect:discussions} for more discussions). 

\begin{figure*}
    \centering
    \includegraphics[width=0.98\textwidth]{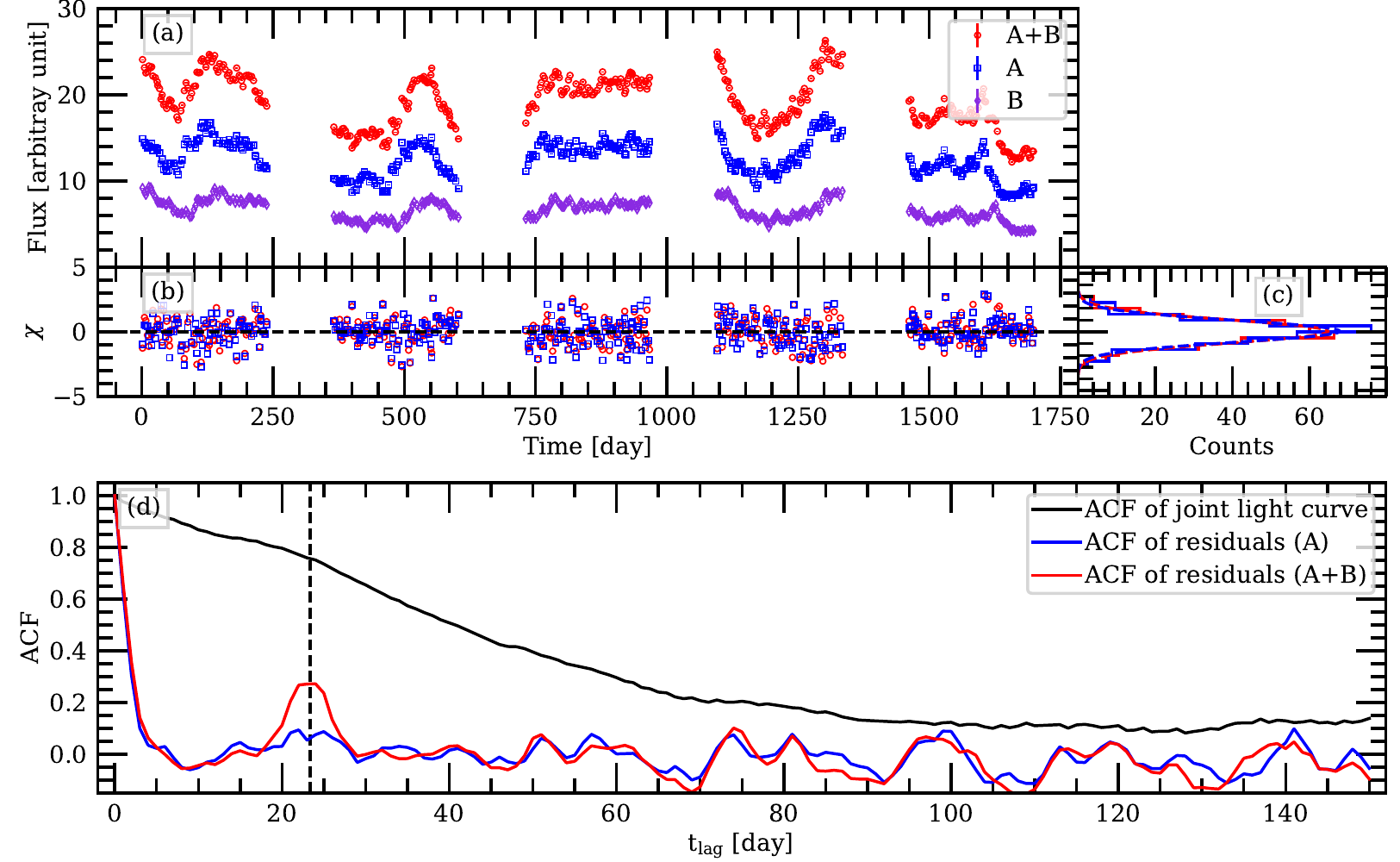}
    \caption{An illustration of the basic idea of our method. Panel (a): Simulated light curves of two images (A and B) in an example strong-lens QSO system from TDC1 and their joint light curve. The relative flux ratio and time delay between A and B are 0.31 and 23.39 days; Panel (b): Normalised residuals from subtracting best-fit DRW models from the joint light curve (red) and the light curve of image A (blue) respectively; Panel (c): Histograms of the normalised residuals of the joint light curve (red) and the light curve of image A (blue), along with Gaussian fits to the histograms (dashed lines) suggesting that the normalised residuals of both fits follow normal distributions; Panel (d): ACFs of the joint light curve (black), the normalised residuals of the joint light curve (red), and normalised residuals of image A (blue). A strong peak at $\sim$23 days is seen in the ACF of the normalised residuals of the joint light curve (vertical dashed line), which is related to the time delay between A and B.}
    \label{fig:test}
\end{figure*}

To demonstrate our idea, we first generate a joint light curve of two lensed QSO images by combining simulated light curves of two individual lensed QSO images in an example lens system selected from the Time Delay Challenge 1 (TDC1). TDC1 light curves are generated by the DRW model. More details are given in the next Section, but a full description of this project can be found in \citet{Dobler15} and \citet{Liao15}. Panel (a) in Figure~\ref{fig:test} shows simulated light curves of two lensed images in the selected lens system and their joint light curve. The relative time delay $t_{\rm delay}$ between these two images is 23.39 days, and the flux ratio of the fainter image to the brighter image (fainter-to-brighter flux ratio hereafter) is 0.31. For this example, no significant excess can be seen in the ACF of the joint light curve around the input time delay, presumably because that signal has been largely diluted by photometric errors and additional variations induced by the microlensing effect. Alternatively, we fit a DRW model to the joint light curve and the light curve of lensed image A respectively following the procedures outlined in \citet{Kelly09}. The DRW model contains two parameters, i.e. the relaxation time scale $\tau$ and the characteristic fluctuation amplitude $\sigma$. The parameter optimisation is done with the ensemble sampler provided by {\tt emcee} \citep{EMCEE}. The best-fit DRW model is generated using $\tau$ and $\sigma$ values inferred from their marginalised posterior probability distribution. Fitting the normalised residuals $\chi$, defined by Equation (13) in \citet{Kelly09}, of both fits by a Gaussian profile, we find that they are well described by a normal distribution, specifically mean=0.04, standard deviation=0.99 for the residuals of the joint light curve and mean=0.05, standard deviation=0.92 for the residuals of image A. It suggests that, from the point view of $\chi^2$ or likelihood alone, a single DRW model can provide a good fit to this joint light curve and the light curve of a single QSO image. However, the normalised residuals of the joint light curve are clearly not uncorrelated as suggested by a strong peak at $\sim$23 days in their ACF (i.e. panel (d)), which matches well with the input time delay of 23.39 days. On the other hand, no strong peak is seen in the ACF of the normalised residuals of image A, although it does have some wiggles. This test confirms our postulation that the ACF of the normalised residuals from a DRW model fit to a light curve can be used for strongly lensed QSO searches. 

\begin{figure}
    \centering
    \includegraphics[width=0.49\textwidth]{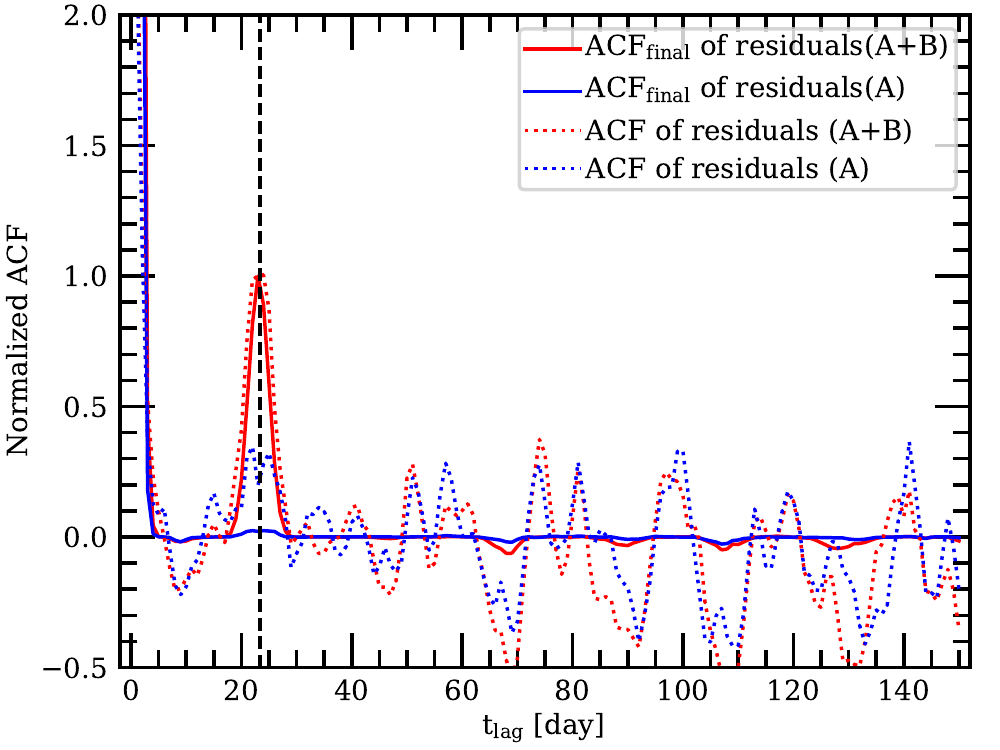}
    \caption{Comparisons of the ACFs of the residuals of the joint light curve (red dotted line) and the light curve of image A (blue dotted line) shown in Figure~\ref{fig:test} with their corresponding ACF$_{\rm final}$ assuming $f_{\rm resampling}=0.98, n_{\rm multiply}=3$ (solid lines). The two dotted lines are rescaled by the ACF value of the residuals of the joint light curve at $t_{\rm lag}=23$ days, while the two solid lines are rescaled by the ACF$_{\rm final}$ value of the residuals of the joint light curve at the same $t_{\rm lag}$. Random peaks seen in the ACF of residuals have been significantly diminished in the ACF$_{\rm final}$ for both the lens and non-lens cases.}
    \label{fig:test2}
\end{figure}

The workflow of our method is thus as follows. For any given system with a measured light curve, a normalised residual curve $\chi (t_i)$ is obtained by subtracting a best-fit DRW model from its light curve. As shown by Figure~\ref{fig:test}, the ACFs of the  normalised residual curve of both lens and non-lens systems can have wiggles at $t_{\rm lag}$ other than $t_{\rm delay}$ primarily due to the quality of light curves (i.e. errors, sampling cadence, length). We therefore compute ACFs of $N$ resampled residual light curves that are generated by randomly selecting some portion, parameterised as $f_{\rm resampling}$, of the time points $t_i$ and assigning to them the same number of normalised residual values that are separately randomly selected and ordered by their original time points. The 1$\sigma$ uncertainty level, $\delta (\rm ACF)$, is constructed by averaging the 16th and 84th percentiles of the $N$ ``resampled'' ACFs at each $t_{\rm lag}$. To further diminish random ``peaks'' in the ACF, we multiply every $n_{\rm multiply}$ resampled ACFs to obtain $N/n_{\rm multiply}$ ``stacked'' ACFs, the average of which, ACF$_{\rm final}$, is taken as the final ACF. By comparing the ACFs of residuals shown in Figure~\ref{fig:test} to their corresponding ACF$_{\rm final}$ assuming $f_{\rm resampling}=0.98, n_{\rm multiply}=3$ (all rescaled), Figure~\ref{fig:test2} demonstrates the benefit of using ACF$_{\rm final}$, in which most of the random peaks seen in the ACFs of residuals in Figure~\ref{fig:test} are significantly if not completely diminished. The choice of $f_{\rm resampling}=0.98, n_{\rm multiply}=3$ here is examined in the next section. We then identify all local maxima (i.e. peaks) in ACF$_{\rm final}$ and define the signal-to-noise ratio (S/N) as ACF$_{\rm final}$/$\delta ({\rm ACF}^{n_{\rm multiply}})$ where $\delta ({\rm ACF}^{n_{\rm multiply}}) \equiv n_{\rm multiply} \times \lvert \text{ACF}_{\rm final}^{1-1/n_{\rm multiply}} \rvert \times \delta (\rm ACF)$. We refer to the local maxima with S/N$\geq$1 as ``hits'', and classify this object as a strong-lens candidate if its total number of hits is between one and five. This upper limit of five is chosen to maintain a reasonable balance between the true positive rate (TPR) and FPR. We note that the local maxima of ACF$_{\rm final}$ are compared to $\delta ({\rm ACF}^{n_{\rm multiply}})$ instead of $\delta (\rm ACF)$ because ACF$_{\rm final}$ has the dimensionality of ACF$^{n_{\rm multiply}}$. We note that linear interpolations are used in the ACF calculations when needed. We have tested that changing to quadratic or cubic spline interpolations leads to lower TPRs and FPRs in general.

\section{Tests with Simulated Light Curves}
\label{sect:test_tdc1}

\subsection{Parameter Configuration}

\begin{figure}
    \centering
    \includegraphics[width=0.49\textwidth]{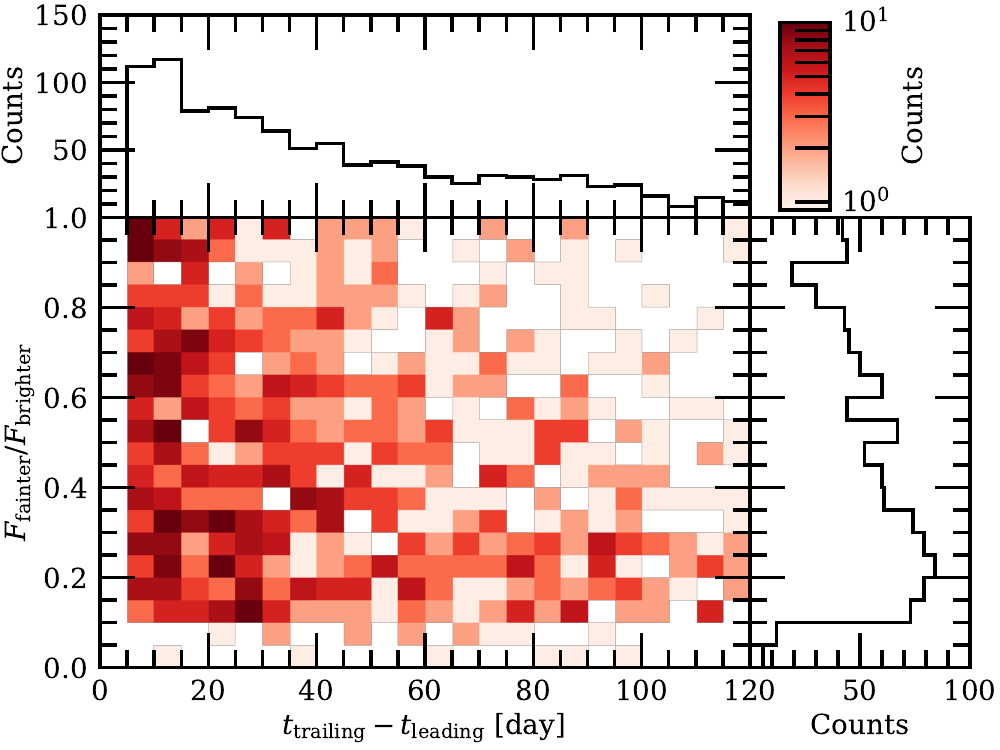}
    \caption{Two-dimensional and one-dimensional distributions of the relative time delay and intrinsic fainter-to-brighter flux ratio (before microlensing is applied) for the 1024 image pairs in TDC1.}
    \label{fig:tdelay_mu_distribution}
\end{figure}

Two main parameters in our method, i.e. $f_{\rm resampling}$ and $n_{\rm multiply}$, need to be configured according to the quality of the input light curves and the required TPR and FPR. We thus employ simulated light curves from TDC1 that were designed to represent realistic light curves for a sample of strongly lensed QSOs. They cover a range of light curve qualities from what has already been achieved by ongoing surveys such as Pan-STARRS1 MDS and ZTF to what will be delivered by the forthcoming time-domain surveys such as LSST. Specifically, TDC1 adopted the mock strongly lensed QSO catalogue in \citet{Oguri10} that was constructed from realistic populations of lensing galaxies and background QSOs. The light curves of background QSOs are generated by the DRW model, with the two DRW parameters randomly drawn from distributions found for a sample of spectroscopically confirmed QSOs \citep{MacLeod10}. Additionally, the microlensing effect and statistical and various systematic errors in epoch photometry (assuming LSST-type observations) have been properly included. In total, TDC1 provides five sets of simulated light curves with different observing parameters (including mean cadence, cadence dispersion, number of observing epochs per year, and total number of observing epochs), known as rung0--4. Each rung contains resolved light curves for the same 1024 independent lensed image pairs in 720 doubly-imaged QSOs (doubles) and 152 quadruply-imaged QSOs (quads) selected from the mock strongly lensed QSO catalogue in \citet{Oguri10}. We note that the four images in any given quad are treated as two independent pairs in TDC1, so only time delays between those two pairs are provided. As shown in Figure~\ref{fig:tdelay_mu_distribution}, the 1024 image pairs in TDC1 have relative time delays between 5 and 120 days (by design) and intrinsic fainter-to-brighter flux ratios (i.e. before microlensing is applied) between 0 and 1. There is some correlation between the relative time delay and intrinsic fainter-to-brighter flux ratio for the 1024 image pairs in the sense that image pairs with longer time delays have lower fainter-to-brighter flux ratios on average. This correlation is understandable as long time delays usually indicate asymmetrical imaging configurations and significant potential differences, which usually lead to significantly different magnifications and hence flux ratios. 

For each rung, we construct a lens data set by directly adding the two light curves (in flux) of each image pair into one joint light curve. The errors of the joint light curve are obtained by adding the flux errors of the two input light curves in quadrature. For this parameter configuration process, we do not consider quads in the lens data set because 1) the TPRs inferred from these TDC1 quads are likely inaccurate as only two of the six putative time delays in each quad are known to us; 2) the fraction of quads among all detectable strongly lensed QSOs is low, $\sim$13 per cent for a wide range of limiting magnitudes $21 \lesssim i \lesssim 27$ \citep[e.g.,][]{Oguri10}. The non-lens data set is constructed to cover two sub-classes, i.e. single QSOs and binary QSOs. The single-QSO sub-class contains 1024 light curves of the brighter images in the 1024 image pairs. The binary-QSO sub-class contains joint light curves for 1024 mock, non-lens QSO pairs, which are constructed by selecting one QSO from each of the two randomly-selected, different image pairs. Multiplets with more than two QSOs are not considered here because QSO multiplets with separations comparable to galaxy-scale, strongly lensed QSOs are extremely rare. In total, the lens data set contains joint light curves for 1024 lens objects, and the non-lens data set contains joint light curves for 2048 non-lens objects.

\begin{table*}
\centering
\begin{tabular}{c c c c c c c c}
\hline
Rung & Cadence & Epochs & TDC1 Success Fraction & ($f_{\rm resampling}$, $n_{\rm multiply}$) & TPR (double) & TPR (quad) & FPR \\
(1) & (2) & (3) & (4) & (5) & (6) & (7) & (8) \\
\hline
0 & $\mathcal{N}(3,1)$ & 80 yr$^{-1} \times \phantom{1}5$ yrs & 33--68 per cent & (0.98, 3) & 20.9 per cent & 27.6 per cent & 10.3 per cent \\ 
1 & $\mathcal{N}(3,1)$ & 40 yr$^{-1} \times 10$ yrs & 22--37 per cent & (0.98, 3) & 17.0 per cent & 17.1 per cent & 8.9 per cent \\ 
2 & $\mathcal{N}(3,0)$ & 40 yr$^{-1} \times \phantom{1}5$ yrs & 18--37 per cent & (0.97, 3) & \phantom{1}9.1 per cent & 16.4 per cent & 9.7 per cent \\ 
3 & $\mathcal{N}(3,1)$ & 40 yr$^{-1} \times \phantom{1}5$ yrs & 19--35 per cent & (0.98, 3) & 10.9 per cent & 17.1 per cent & 12.2 per cent \\ 
4 & $\mathcal{N}(6,1)$ & 20 yr$^{-1} \times 10$ yrs & 16--35 per cent & (0.99, 5) & \phantom{1}7.0 per cent & \phantom{1}9.2 per cent & 9.3 per cent \\
\hline
\end{tabular}
\caption{Performances of our method on the five rungs of TDC1. Column (2) indicates the cadence distribution of each rung (in days) with $\mathcal{N}(\mu,\sigma)$ representing a normal distribution of mean $\mu$ and standard deviation $\sigma$. Column (3) gives the total number of epochs in the format of a product between number of epochs per year and number of years. Column (4) shows the range of fraction of image pairs for which time delays can be successfully determined from their light curves by nine different algorithms examined in TDC1. Column (5) gives the reference configuration of the two parameters in our method. Column (6)--(8) give the TPR for doubles, TPR for quads, and FPR by applying our method with the reference configuration to the TDC1 light curves.}
\label{tab:TDC1_performance}
\end{table*}
\begin{figure*}
    \centering
    \includegraphics[width=0.48\textwidth]{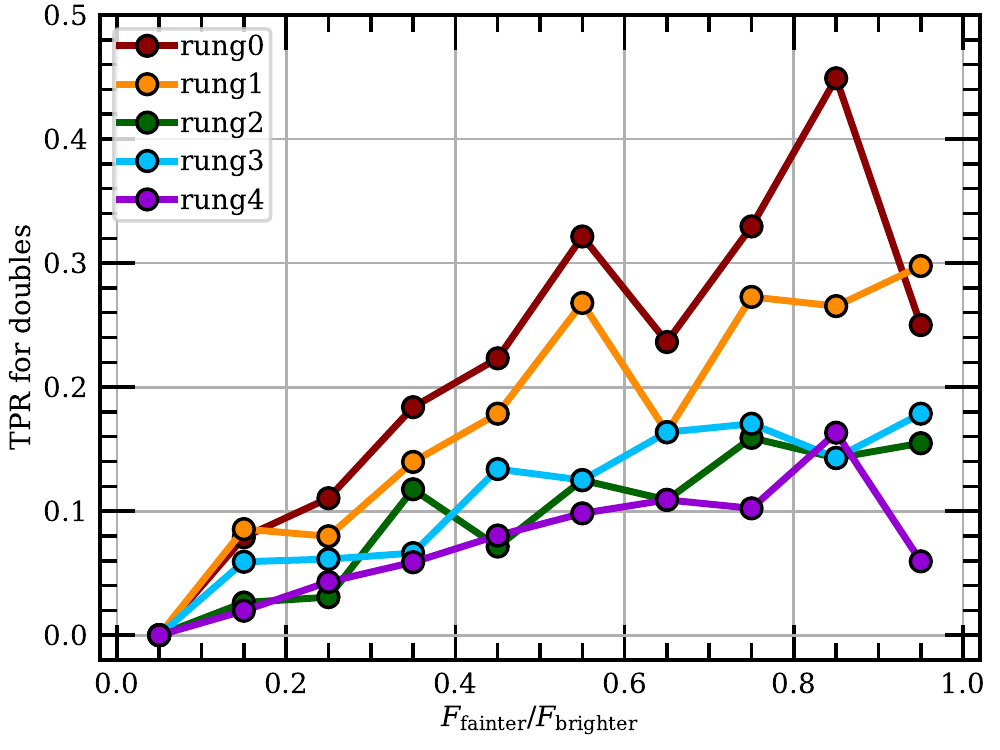}
    \includegraphics[width=0.48\textwidth]{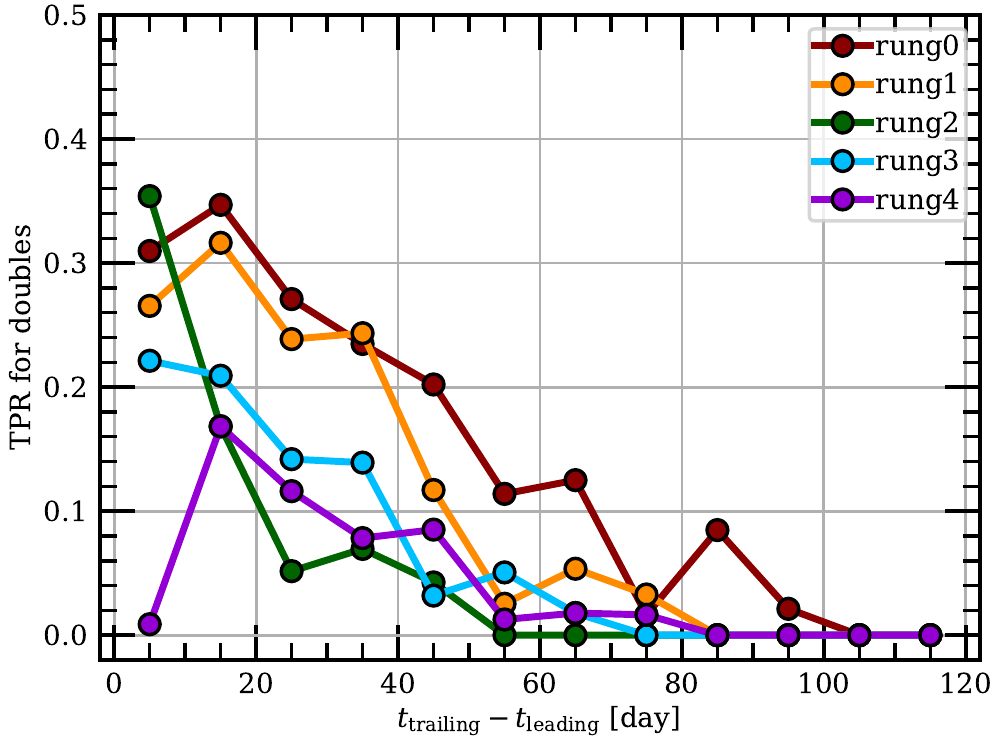}
    \caption{TPRs for doubles as a function of the intrinsic flux ratio (\emph{left}) and relative time delay (\emph{right}) for the five TDC1 rungs. }
    \label{fig:performance}
\end{figure*}

To determine the optimal values of $f_{\rm resampling}$ and $n_{\rm multiply}$ for each rung, we perform a grid search with $f_{\rm resampling}$ from 0.90 to 0.99 with a step size of 0.01 and $n_{\rm multiply}$ from 1 to 10 with a step size of 1. The $t_{\rm lag}$ baseline for the ACF computation is chosen to be from 0 to 150 days with a step size of 1 day. For each $n_{\rm multiply}$, the number of resampled residual curves $N$ is set to $100 \times n_{\rm multiply}$. At each grid point, we identify all the lens candidates in the lens and non-lens data sets using the method and criteria explained above. The FPR is simply the fraction of objects in the non-lens data set that are classified as lens candidates. Among the classified lens candidates in the lens data set, only those that have at least one hit located at a time lag that is within $\pm1$ day of the input time delay are considered as successfully recovered. The overall TPR is therefore defined as the number of recovered lens candidates from the lens data set divided by 1024. Our choice of using the recovered lens candidates for the TPR computation ensures that the TPRs quoted in this work represents our method's ability of discovering strongly lensed QSOs for which time delays can already be estimated from input light curves.

\subsection{Performances on TDC1}

Generally speaking, we find that both TPR and FPR increase with $f_{\rm resampling}$ at a fixed $n_{\rm multiply}$ because more structures in the ACF are preserved when more data points are used. At a fixed $f_{\rm resampling}$, TPR and FPR generally decrease with $n_{\rm multiply}$ because a peak will be classified as a hit only if its ACF$^{n_{\rm multiply}} / \delta ({\rm ACF}^{n_{\rm multiply}}) \equiv \frac{1}{n_{\rm multiply}} \frac{\text{ACF}}{\delta(\text{ACF})}$ ratio is at least 1. Increasing $n_{\rm multiply}$ will progressively reduce this ratio and leads to less hits. Furthermore, multiplying more resampled ACFs tends to suppress random peaks more aggressively. The FPRs for the two non-lens sub-classes are generally comparable, with no noticeable systematic differences. 

In this work, we consider a FPR of $\lesssim$10 per cent as acceptable, and choose the combination of $f_{\rm resampling}$ and $n_{\rm multiply}$ that delivers the highest overall TPR while satisfying FPR $\lesssim$10 per cent as a reference configuration for each rung. We find that the so-obtained reference configuration is $f_{\rm resampling}=0.98$ and $n_{\rm multiply}=3$ for rung0, rung1 and rung3. The reference configurations are $f_{\rm resampling}=0.97$ and $n_{\rm multiply}=3$ for rung2 and $f_{\rm resampling}=0.99$ and $n_{\rm multiply}=5$ for rung4. In addition, we construct another quad lens data set for each rung by combining all the four light curves of a given quad into a joint light curve, and consider a quad to be recovered if it has at most five hits in total including at least one hit located at a time lag that is within $\pm1$ day of either of the two input time delays known to us. Each quad lens data set contains 152 light curves. We note that the TPRs for quads in this work are expected to be underestimated because 1) the upper limit of five hits in total is presumably too restrictive for quads, which have up to six different time delays that could leave imprints in the ACF; 2) quads with hits only around the other time delays that are not given by TDC1 will be missed. 

Adopting the reference configurations, we find that the TPRs for doubles and quads are approximately 7--21 per cent and 9--28 per cent respectively (Table~\ref{tab:TDC1_performance}). TPRs for quads are generally higher than for doubles, which is understandable because quads potentially have more peaks in the ACFs and are therefore more likely to be picked out by our method. It also suggests that fitting the joint light curve of a quad with a single DRW model still substantially preserves correlations at different characteristic time scales. The FPRs are in the range of 9--12 per cent. We have tested that the reported TPRs for the five rungs are insensitive to the uncertainties in the DRW fit. Among the five rungs, rung0 has the highest TPRs, which is expected given its densest sampling, i.e. on average 80 epochs per year with a mean cadence of 3 days. Likewise, rung4 has the lowest overall TPRs because it has the longest mean cadence (i.e. 6 days). 
The results also suggest that the TPRs will be higher when light curves are more densely sampled (by comparing rung0 with rung1) or light curves span a longer time baseline (by comparing rung1 with rung3). Moreover, we bin objects in the lens data set of each rung by their flux ratios or time delays, and show the specific TPR for doubles in each bin in Figure~\ref{fig:performance}. We find that the TPR for doubles generally increases with the flux ratio, reaching up to almost 50 per cent. This is expected given that the level of ACF excess is directly related to the flux ratio (Equation~(\ref{eq:ACF_joint})). The TPR is also found to decrease with the time delay, which could be related to the flux ratio-time delay correlation for the sample considered. Additionally, fewer pairs of data points are available for the ACF computation at larger $t_{\rm lag}$, which could lead to stronger variations and/or weaker peaks in the ACF, and consequently lower TPRs. As a result, strongly lensed QSOs discovered by this method will be incomplete and biased, and can not be directly used in certain applications that require statistical samples of strongly lensed QSOs, for example, constraining cosmological parameters with the lensing probability \citep[e.g.,][]{Mitchell05, Chae07, Oguri12}, inferring the lens mass profiles and the Hubble constant with the distribution of time delays \citep[e.g.,][]{Oguri02, Huterer04, Oguri07}, etc.

In fact, 10 different algorithms from seven teams that participated in TDC1 have been applied to the five rungs to measure time delays between image pairs from resolved light curves. These are among the current best algorithms for the time-delay measurement. \citet{Liao15} examined the success fraction, defined as the fraction of image pairs for which time delays were successfully determined, of each algorithm on each rung. In Table~\ref{tab:TDC1_performance}, we show the range of the success fraction for each rung achieved by nine of the 10 algorithms as one algorithm has very low success fractions of 2--4 per cent in all the five rungs. The overall success fraction is between 16 per cent and 68 per cent, and it is significantly higher for rung0 than for the other four rungs, which have similar success fractions. As pointed out by \citet{Liao15}, the 10 algorithms all failed for 20--40 per cent of the light curves (depending on which rung) as those light curves simply do not contain enough features for any of the algorithms to succeed. The rest of the failures vary from algorithm to algorithm, suggesting that different algorithms also rely on different features. Although details about which algorithm failed on which systems are not known to us for a direct comparison, we expect our method to have difficulties on those failed light curves as well. In that sense, our TPRs should be interpreted within the context of the TDC1 success fractions, even though the light curves are analysed differently between our method and the time-delay measurement algorithms. We therefore define an effective TPR as the ratio of our TPR to the average of the lower and upper bounds of the TDC1 success fractions. The effective TPRs are 28--58 per cent for doubles and 36--60 per cent for quads. It also suggests that presumably all strongly lensed QSOs discovered by our method will be guaranteed to have robust time-delay measurements from existing algorithms, and therefore can be used for cosmological studies. 

\section{Tests with Observed Light Curves of known lenses}
\label{sect:test_real}

Our method relies on the assumption that the input light curves of single QSOs can be described by a DRW model, and so far we have only tested it with TDC1 simulated light curves that were generated using the DRW model. Although many studies on large samples of QSOs show that the DRW model indeed provides a good description of the optical variability of QSOs \citep[e.g.,][]{Kelly09, Kozlowski10, MacLeod10, Zu13}, it is still worth testing our method on some real data. The COSMOGRAIL project has been monitoring dozens of known strongly lensed QSOs for more than a decade \citep{Eigenbrod05}, and provides the current largest light curve database for strongly lensed QSOs. In fact, time delays determined from the COSMOGRAIL resolved/deblended light curves have enabled independent measurements of the Hubble constant \citep[e.g.,][]{Bonvin17, Chen19, Wong20}. We therefore choose to test our method on COSMOGRAIL light curves. 

\begin{figure}
    \centering
    \includegraphics[width=0.48\textwidth]{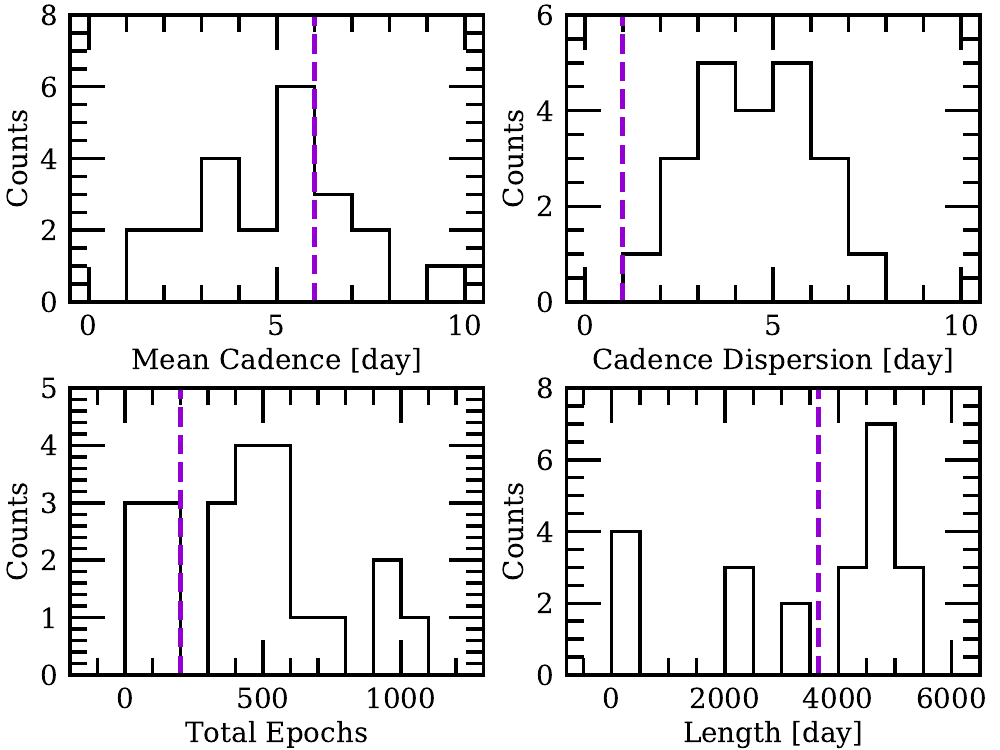}
    \caption{Properties of the COSMOGRAIL light curves for the 22 lenses studied here. From top to down, it shows the histograms of mean cadence, dispersion of the cadence, total number of epochs, and the total length of the light curves. The dashed violet lines indicate the same parameters for rung4 in TDC1.}
    \label{fig:cosmograil_lcs_statistics}
\end{figure}
\begin{table} 
\begin{threeparttable} 
\centering 
\caption{\label{tb:known_lens_recovery} Measured flux ratios and time delays and hits detected by our method (ordered by their S/Ns) for the 5 recovered, known strongly lensed QSOs. For each image pair, the fainter-to-brighter flux ratio is approximated by the ratio of the median fluxes of that pair. The first two are doubles, and the remaining three are quads. HE\,0435$-$1223 appears twice because two sets of COSMOGRAIL light curves for this system are provided, one set covers $\approx$7 years while the other set adds another 8-year data on top of the previous set. Images A and B in HE\,0230$-$2130 are not resolved in the data, so their fluxes are combined as for one single virtual image A$^{\prime}$ (see \citet{Millon20} for details). References for the discoveries and time-delay measurements are given at the bottom of the table. }
\begin{tabular} {l l l l} 
\hline 
\hline 
System & Flux ratio & Time delay [day] & Hits [day] \\ 
\hline 
Q\,J0158$-$4325$^{1}$ & 0.37 & $-22.7 \pm 3.6 ^{a}$    & (19) \\ 
\hline
HS\,2209$+$1914$^{2}$ & 0.82 & $20.0 \pm 5.0 ^{b}$     & (15, 12) \\ 
\hline 
HE\,0230$-$2130$^{3}$ & 0.24 & A$^{\prime}$C: $15.7 \pm 3.9 ^{a}$     & (17) \\ 
\hline
HE\,0435$-$1223$^{4}$ & 0.57 & AB: $-8.4 \pm 2.1 ^{c}$ & (10, 8, 13) \\ 
 & 0.59 & AC: $-0.6 \pm 2.3 ^{c}$ & \\
 & 0.50 & AD: $-14.9 \pm 2.1 ^{c}$ & \\
 & 0.97 & BC: $7.8 \pm 0.8 ^{c}$ & \\
 & 0.87 & BD: $-6.5 \pm 0.7 ^{c}$ & \\
 & 0.84 & CD: $-14.3 \pm 0.8 ^{c}$ & \\
\hline
HE\,0435$-$1223$^{4}$ & 0.58 & AB: $-9.0\pm 0.8 ^{a}$   & (8, 10, 13, 23) \\ 
 & 0.62 & AC: $-0.8 \pm 0.7 ^{a}$ & \\
 & 0.50 & AD: $-13.8 \pm 0.8 ^{a}$ & \\
 & 0.95 & BC: $7.8 \pm 0.9 ^{a}$ & \\
 & 0.86 & BD: $-5.4 \pm 0.8 ^{a}$ & \\
 & 0.81 & CD: $-13.2 \pm 0.8 ^{a}$ & \\
\hline 
\hline 
\end{tabular} 
$^1$ \citet{Morgan99}; $^2$ \citet{Hagen99}; $^3$ \citet{Wisotzki99}; $^4$\citet{Wisotzki02}. \\
$^a$ \citet{Millon20}; $^b$ \citet{Eulaers13}; $^c$ \citet{Courbin11}.
\end{threeparttable} 
\end{table} 

At the time of writing, 36 sets of COSMOGRAIL light curves for 21 (20 unique) doubles and 15 (9 unique) quads have been released\footnote{\url{https://obswww.unige.ch/~millon/d3cs/COSMOGRAIL_public/code.php}}, as some lenses have been observed by more than one telescopes or across different time baselines. For our purpose, we can treat them as 36 independent lenses. Among them, time delays for four doubles and one quad can not be determined from those light curves. Time-delay measurements for the rest are provided in \citet{Courbin11}, \citet{Eulaers13}, \citet{Tewes13}, \citet{Kumar13}, \citet{Courbin18}, \citet{Bonvin18}, \citet{Bonvin19}, and \citet{Millon20}. To make a fair comparison, we further discard seven doubles and two quads for which the measured time delays are either less than 5 days or larger than 120 days within 1$\sigma$ uncertainties. 

For each of the remaining 10 doubles and 12 quads, we construct a joint light curve from the resolved/deblended COSMOGRAIL light curves. The properties of the COSMOGRAIL light curves for these 22 lenses are quite heterogeneous. The mean cadence ranges from 1.65 days to 9.69 days with a median cadence dispersion of 4.66 days. The length of the light curves ranges from 178 days to 5491 days, and the total number of epochs ranges from 48 to 1002 (Figure~\ref{fig:cosmograil_lcs_statistics}). Comparing to TDC1 light curves, the quality of the COSMOGRAIL light curves is generally poorer, and is most closest to rung4 in TDC1. We therefore choose the configuration for rung4, i.e. $f_{\rm resampling}=0.99$ and $n_{\rm multiply}=5$, and apply our method to the 22 joint light curves. A lens is considered to be recovered if it has at most five hits including at least one hit that is within $\pm \sqrt{1+\Delta t_{\rm delay}^2}$ of the absolute value of the measured time delay (for doubles) or any of the measured time delays (for quads), where $\Delta t_{\rm delay}$ is the 1$\sigma$ uncertainty of that measured time delay. 

Our method successfully recovers two doubles and three quads according to the above standard, i.e. 20 per cent of doubles and 25 per cent of quads examined. The measured time delays and hits detected by our method for the five recovered systems are shown in Table~\ref{tb:known_lens_recovery}. Despite the small sample size, the TPRs in this test are comparable to the effective TPRs for rung4, i.e. 28 per cent and 36 per cent. This test demonstrates that our method can work as expected with real data. 

The encouraging matches between the measured time delays and locations of hits identified by our method for the known lenses suggest that it also has the potential of directly measuring time delays of strongly lensed QSOs, especially when only unresolved light curves are available. However, due to the nature of the ACF analysis, only the absolute values of the time delays can be inferred by our method. It is therefore not straightforward to determine which is the leading image for doubles or what are the pairwise time delays for quads, which will substantially limit further time delay-related applications.

\section{Discussion}
\label{sect:discussions}

The performance of our method is affected by several free parameters including $f_{\rm resampling}$, $n_{\rm multiply}$ and the maximum number of hits allowed. Different combinations of these parameters yield different TPRs and FPRs. In this work, we set the maximum number of hits allowed to be five and show TPRs and FPRs for TDC1 light curves with reference configurations that achieve the highest TPRs at FPRs$\lesssim 10$ per cent as an example. In practise, these free parameters should be adjusted according to the quality of the input light curves and the specific demands on the TPR and FPR. For example, changing $(f_{\rm resampling}, n_{\rm multiply})$ to $(0.99, 5)$ for rung0, the FPR drops substantially from 10.3 per cent to 3.6 per cent while the TPR for doubles decreases only moderately from 20.9 per cent to 14.4 per cent. If the maximum number of hits allowed is increased from five to six, one more double examined in Section~\ref{sect:test_real} will be recovered. 

In fact, using the TDC1 light curves potentially overestimates the FPRs of our methods for light curves with properties similar to the five TDC1 rungs. The TDC1 light curve of any QSO image is generated under the presence of a lensing galaxy that almost certainly introduces extra variations caused by the microlensing effect, considering that the microlensing optical depth at the position of a strongly lensed QSO image is of order unity \citep[e.g.,][]{Chang79, Narayan96, Schmidt10}. On the other hand, for true contaminants such as unlensed QSOs and QSO multiplets that generally do not have massive structures in their close proximity, their light curves should be barely affected by microlensing because the microlensing optical depth due to the Milky Way or other compact mass structures because of chance alignment is on the order of $\sim 10^{-6}$ \citep[e.g.,][]{Belokurov04, Tisserand07, Wang11, Mroz19}. In principle, extra variations due to microlensing in the TDC1 light curves could introduce extra correlations at non-zero $t_{\rm lag}$ (e.g., \citealt{Sliusar15}; \citealt{Fedorova16}, see also Figure 3 in \citealt{Liao15}), which will artificially raise the FPRs. Nevertheless, this overestimation is expected to be small because the correlation time scales due to microlensing are generally larger than the time lags examined here, i.e. 0--150 days. 
%In fact, many false positives happen at t_lag ~ 10 days 

We have so far considered single and binary QSOs as contaminants. In reality, the population of point sources on the sky is dominated by stars. However, as a pre-selection on the level of variability is usually expected when applying our method, the vast majority of stars will hence be removed. Other plausible contaminants include variable stars and star pairs and variable QSO-star superpositions. As estimated by \citet{Kochanek06}, the overall surface densities of variable star pairs and QSO-star superpositions in the magnitude range of 19--24 $V$ mag are generally lower than lensed QSOs at high Galactic latitudes ($\lvert b \rvert \gtrsim 20^{\circ}$). Regarding variable stars that are more common than lensed QSOs, how well the DRW model fit will be is unclear to us, and so is their FPR. Nonetheless, morphology, multi-band photometry, and astrometry are generally available, from either the time-domain survey itself or other existing surveys, and can be used to effectively eliminate variable stars \citep[e.g.,][]{Richards02, Stern05, Wu10, Gaia16, Lemon18, Shu19}. We are not particularly concerned about contamination from QSO multiplets with small separations ($\lesssim 3^{\prime \prime}$) because 1) they are less abundant than strongly lensed QSOs \citep[e.g.,][]{Mortlock99, Hennawi06}; 2) small-separation QSOs multiplets are also important events that can provide insight into structure growth, galaxy evolution, cosmology, and general relativity \citep[e.g.,][]{Volonteri03, Kormendy09, Centrella10, Colpi11}. It is therefore expected that false positives directly from our method are dominated by single QSOs, which can be largely eliminated afterwards by examining the morphology (i.e. extendedness). 

Although this work uses the DRW model to assess the variability of single QSOs, our method can be straightforwardly coupled with other models that can also provide good descriptions of the QSO variability. For example, studies have found that QSOs can exhibit a range of power spectrum density slopes different from what the DRW model suggests, especially towards shorter timescales ($\sim$days)  \citep[e.g.,][]{Mushotzky11, Zu13, Kasliwal15, Caplar17, Aranzana18}. Recently, \citet{Sun20} proposed that the Corona-Heated Accretion-disk Reprocessing model can be used to describe the optical variability of QSOs. 

Additionally, our method can be easily extended to work with multi-filter light curves, which is expected to further improve the performance as long as the QSOs show detectable variability in the filters considered. Fundamentally, our method exploits the correlated variations between lensed QSO images induced by the strong-lensing effect, which, in itself, is achromatic in vacuum. In principle, effects such as differential lensing, microlensing, and plasma lensing can introduce chromatic alterations to the arrival time of individual lensed QSO images. However, the changes in the relative time delays between different filters due to those effects are generally insignificant for the analysis in this work \citep[e.g.,][]{Er14, Kogan17, Liao20}. As a result, combining the ACFs of the residuals in different filters (for example, by multiplication) is expected to enhance the true signal around the time lag that corresponds to the time delay and depress random peaks in the ACFs of individual filters at the same time. Moreover, it is feasible to fit light curves in different filters with different models that best describe the QSO variability in the filters considered. 

Our method is well suited for discovering galaxy-scale, strongly lensed QSOs in time series data from current and forthcoming large-scale time-domain surveys such as Pan-STARRS1 MDS, ZTF, and LSST. In particular, our method only requires unresolved photometry instead of resolved photometry, which can be challenging to accurately construct because the typical angular resolution of the current and forthcoming time-domain surveys is comparable to the typical image separations in galaxy-scale strongly lensed QSOs. Even if resolved photometry can be obtained, it is computationally cheaper to cope with unresolved photometry than resolved photometry for the purpose of discovery. 

In terms of expected performances, Pan-STARRS1 MDS was a 4-year time-domain survey with a nominal cadence of 3 days in $griz$ filters and observing season of 6--8 months \citep{PS1}. ZTF is an ongoing 3-year time-domain survey with two components. The ZTF Northern Sky Survey has a nominal cadence of 3 days in $gr$ filters, and the ZTF Galactic Plane Survey has a nominal cadence of 1 day in $gr$ filters. The observing season for the two surveys in ZTF is 6--8 months \citep{ZTF}. Therefore, the quality of light curves from Pan-STARRS1 MDS and ZTF are expected to be mostly similar to rung0. LSST is a scheduled 10-year time-domain survey. Although the observing strategy of LSST has not been fully determined yet, rung1--4 were actually designed by TDC1 to match some of LSST temporal sampling options under consideration \citep{Liao15, Marshall17, Huber19}. 

More importantly, our method is particularly useful for discovering strongly lensed QSOs with small image separations that can be challenging for imaging-based methods that tend to bias towards large image separations, although it can be improved when assisted with resolved astrometry, for example, from \emph{Gaia} \citep[e.g.,][]{Lemon17, Lemon18, Lemon19}. Small-separation strongly lensed QSOs are of particular importance scientifically. For instance, due to the nature of gravitational lensing, smaller image separations usually happen when the lensing galaxy is less massive and/or at higher redshift. In general, analyses of strong lenses can provide observational constraints on the distributions of dark matter, stars, and interstellar medium in galaxies at cosmological distances, which have deepened our understanding of various physical processes that regulate galaxy formation and evolution \citep[e.g.,][]{Hewett94, Kanekar05, Treu06, Bolton08, Suyu09, Auger10, Bolton12, Sonnenfeld13, Oguri14, Shu15, Lipnicky18, Li18, Kulkarni19}. Small-separation lenses will enable such studies in lower-mass and higher-redshift galaxies that are much less explored within the current strong-lens samples, and help to build a more complete picture on galaxy formation and evolution. Additionally, more inner regions of the lensing galaxies are probed in smaller-separations lenses, where the total masses are more dominated by stars. At the same time, the microlensing optical depth increases quickly towards galaxy centre. Thus more stringent constraints on the stellar IMF can be obtained from smaller-separations lenses using both strong lensing and microlensing effects \citep[e.g.,][]{Treu10, Auger10a, Ferreras10, Brewer14, Spiniello15, Jimenez-Vicente19}. 

%Pan-STARRS1 Medium Deep Survey (6--7 day cadence, Scolnic et al. (2018)), periodically varying quasars, 

%SMBHBs (Liu T. et al., 2016). 

%Hirv et al. 2007 discussed measuring time delays from unresolved photometry. 

%color light curve drw fit fixing time scale

%HSC same pipeline as LSST

%biased towards doubles

\section{Conclusions}
\label{sect:conclusion}

In this work, we present a new method of discovering strongly lensed QSO systems using their joint light curves. We show that by subtracting a best-fit model that can describe variations of a single QSO from the joint light curve of a strongly lensed QSO, the autocorrelation function (ACF) of the resulting residuals will exhibit excesses at time lags that correspond to the time delays due to the correlated variations between lensed QSO images induced by the strong lensing effect, which can therefore be used as a strong-lens indicator. We further introduce several adjustable parameters to enhance the ACF excesses due to strong lensing and at the same time suppress structures caused by other strong lensing-unrelated effects. 

We test our method on five rungs of simulated single-filter light curves from TDC1 with a range of qualities that were designed to match the qualities of realistic light curves expected from current and forthcoming time-domain surveys. We find that the parameters in our method can affect its performance, and therefore need to be adjusted according to the quality of the input light curves and the actual demand on the TPR and FPR. In this work, we specifically configure the parameters for each TDC1 rung such that the TPR reaches the highest while the FPR is $\lesssim$10 per cent. Under these reference parameter configurations, we achieve overall TPRs of 7--21 per cent for doubles and 9--23 per cent for quads while maintain $\lesssim$10 per cent FPRs. We find that the TPRs for doubles increase with the flux ratio between the fainter and brighter lensed images, and reach up to almost 50 per cent for rung0 that has the best-quality light curves among the five rungs. We also detect an anti-correlation between the TPRs for doubles and the relative time delay, which we think is primarily driven by the correlation between TPRs and flux ratios and/or limited lengths of light curves. Considering that a substantial fraction of TDC1 light curves (32--84 per cent depending on the rungs) do not contain enough features for some of the current best algorithms to robustly derive time delays from, we compute the effective TPRs of our method with the reference parameter configurations to be 28--58 per cent for doubles and 36--60 per cent for quads. 

In addition, we apply our method to observed light curves collected by the COSMOGRAIL project of 22 known strongly lensed QSOs (10 doubles and 12 quads) with measured time delays, and successfully recover two (or 20 per cent) doubles and three (or 25 per cent) quads using a reference configuration of parameters. The quality of the COSMOGRAIL light curves for these lenses is generally poorer than TDC1 light curves, and is most closest to rung4. The TPRs on observed light curves are therefore comparable to the effective TPRs for rung4, demonstrating that our method can work as expected with real data. 

Our method is therefore well suited for discovering galaxy-scale, strongly lensed QSOs from existing, ongoing, and forthcoming large-scale time domain surveys such as Pan-STARRS1 MDS, ZTF, and LSST. It is also straightforward to extend our method to work with multi-filter light curves provided by those surveys, which is expected to further reduce the FPR and increase the TPR. Moreover, our method is particularly capable of discovering small-separation strongly lensed QSOs that can be challenging for traditional image-based methods, which will help in gaining a more complete understanding of galaxy formation and evolution and more stringent constraints on the stellar IMF. 

\section*{Acknowledgements}

We thank Dr. Kai Liao for sharing the TDC1 light curves, and Drs. Matthew Auger and Brandon Kelly for helpful discussions. We thank the anonymous referee for constructive comments that improved the quality of the paper. Yiping Shu acknowledges support from the Max Planck Society and the Alexander von Humboldt Foundation in the framework of the Max Planck-Humboldt Research Award endowed by the Federal Ministry of Education and Research.

\section*{Data availability}

All data used in this work are public. TDC1 light curves are available in \citet{Liao15}, and COSMOGRAIL light curves are avialable at \url{https://obswww.unige.ch/~millon/d3cs/COSMOGRAIL_public/code.php}.

%%%%%%%%%%%%%%%%%%%%%%%%%%%%%%%%%%%%%%%%%%%%%%%%%%



%%%%%%%%%%%%%%%%%%%%%%%%%%%%%%%%%%%%%%%%%%%%%%%%%%

%%%%%%%%%%%%%%%%% APPENDICES %%%%%%%%%%%%%%%%%%%%%

%\appendix



%%%%%%%%%%%%%%%%%%%%%%%%%%%%%%%%%%%%%%%%%%%%%%%%%%

% Don't change these lines
\bsp	% typesetting comment
\label{lastpage}
\end{document}